\input{psfig}
\documentclass[oldversion,referee]{aa}
\input epsf
\input psfig.sty
\newcommand{\be}{\begin{equation}}
\newcommand{\ee}{\end{equation}}
\newcommand{\bea}{\begin{eqnarray}}
\newcommand{\eea}{\end{eqnarray}}
\begin{document}
\title{
On the nature of gravitational forces
} 
\author{A. Del Popolo\inst{1,2}
}
\titlerunning{On the nature of gravitational forces}
\authorrunning{A. Del Popolo}
\date{}
\offprints{A. Del Popolo, E-mail:antonino.delpopolo@boun.edu.tr}
\institute{
$1$ Dipartimento di Matematica, Universit\`{a} Statale di Bergamo,
  via dei Caniana, 2,  24127, Bergamo, ITALY\\
$2$ Istanbul Technical University, Ayazaga Campus,  Faculty of Science and Letters,  34469 Maslak/ISTANBUL, Turkey\\
$^2$ Bo$\breve{g}azi$\c{c}i University, Physics Department,
     80815 Bebek, Istanbul, Turkey\\}


\abstract{
In this paper I show how the statistics of the gravitational field is changed when the system is characterized by
a non-uniform distribution of particles.
I show how the distribution functions $ W({\bf F}$) and $ W({\bf F},d{\bf F}/dt)$, giving the joint
probability that a test particle is subject to a force {\bf F} and an associated rate of change of 
{\bf F} given by d{\bf F}/dt, are modified by inhomogeneity. Then I calculate the first moment of d{\bf F}/dt 
to study the effects of inhomogenity on dynamical friction.
Finally I test, 
by N-Body simulations, that the theoretical W(F) and d{\bf F}/dt describes correctly 
the experimental data and I find that the stochastic force distribution obtained for 
the evolved system is in good agreement with theory.
Moreover, I find that in an inhomogeneous background the friction force is actually enhanced relative to the
homogeneous case.
}


\maketitle

\section{Introduction}

\noindent
Several authors have stressed the importance of stochastic forces and in particular dynamical friction
in determining the observed properties of clusters of galaxies (White 1976; Kashlinsky 1986,1987)
while others studied the role of dynamical friction in the orbital decay of a satellite moving around a galaxy or in the merging
scenario (Bontekoe \& van Albada 1987; Seguin \& Dupraz 1996; Dominguez-Tenreiro \& Gomez-Flechoso 1998). 
The study of stochastic forces is not only important 
in the framework of galaxy formation picture, 
but it is also important for the study of particular aspects of the evolution of a number of astronomical
systems, such as galactic nuclei, cD galaxies in rich galaxy clusters.

The study of the statistics of the fluctuating gravitational force in infinite 
homogeneous systems was pioneered by Chandrasekhar \& Von Neumann in two classical papers 
(Chandrasekhar \& Von Neumann 1942, 1943 hereafter CN43) and in several other papers by Chandrasekhar 
(1941, 1943a, 1943b, 1943c, 1943d, 1943e, 1944a and 1944b). The analysis of
the fluctuating gravitational field, developed by the authors, was
formulated by means of a statistical treatment. 
Two distributions are fundamental for the description of the fluctuating 
gravitational field: \\
1) $ W({\bf F})$ which gives the probability that a test star
is subject to a force ${\bf F}$ in the range ${\bf F}$, ${\bf F}$+
d${\bf F}$. $ W({\bf F})$, known as Holtsmark's law (Holtsmark 1919), in the 
case of a homogeneous distribution of the stars, gives information only 
on the number of stars experiencing a given force but it does not 
describe some fundamental features of the fluctuations in 
the gravitational field such as the {\it speed of the fluctuations} and 
the dynamical friction.\\ 
2) $ W({\bf F},{\bf f})$ which gives the
joint probability that the star experiences a force {\bf F} and a rate 
of change {\bf f}, where $ {\bf f} = d{\bf F}/dt $.
The features of gravitational field that the $ W({\bf F})$ does not describe are 
described by this second distribution $ W({\bf F},{\bf f})$.
Hence, for the definition of the speed of fluctuations and of the 
dynamical friction one must determine the distribution $ W({\bf F},{\bf f})$.\\
The function $ W({\bf F}) $ for a homogeneous 
system was obtained for the first time by Chandrasekhar \& von Neumann (1942) 
under the hypotheses that the stars are distributed 
with uniform density in a spherical system,  
that there are no correlations 
and that $\frac{N}{R^{3}} = constant$ when 
$ N \rightarrow \infty$ and $ R \rightarrow \infty$, where N is the 
total number of stars and R the radius of the stellar system. They also 
showed that the force probability distribution is given by 
Holtsmark's law.\footnote{Historically Holtsmark's 
law was calculated to obtain the 
probability of a given electric field strength at a point in a gas 
composed of ions.} 
Spherical symmetry is used so that at the 
center of the system $ {\bf F}_{pot} =0$ and 
$ {\bf F}_{tot} ={\bf F}_{stoch} $, 
while the absence of correlation is required in order to be able to decompose 
the gravitational field into a mean and a stochastic component.\\   
Chandrasekhar's theory (and in particular his
classical formula (see Chandrasekhar 1943b))
is widely employed to quantify dynamical friction
in a variety of situations, even if
the theory developed is based on the
hypothesis that the stars are distributed uniformly and 
it is well known that in stellar systems, the stars are not 
uniformly distributed,
(Elson et al. 1987; Wybo \& Dejonghe 1996; Zwart et al. 1997) 
in galactic systems as well, the galaxies are not uniformly distributed 
(Peebles 1980; Bahcall \& Soneira 1983; Sarazin 1988; Liddle, \& Lyth 1993; 
White et al. 1993; Strauss \& Willick 1995).
It is evident
that an analysis of dynamical friction taking account of the
inhomogeneity of astronomical systems can provide a more realistic
representation of the evolution of these systems.
Moreover from a pure theoretical standpoint we expect that inhomogeneity affects all the aspects of the fluctuating 
gravitational field (Antonuccio \& Colafrancesco 1994).
Firstly, the Holtsmark distribution is no longer correct 
for inhomogeneous systems. For these systems, as shown by Kandrup 
(1980a, 1980b, 1983), the Holtsmark distribution must be substituted 
with a generalized form of the Holtsmark distribution
characterized by a shift of $W({\bf F})$
towards larger forces when inhomogeneity increases. This result 
was already suggested by the numerical simulations of Ahmad \& Cohen 
(1973, 1974). 
Hence when the inhomogeneity increases
the probability that a test particle experiences a
large force increases, secondly, 
$ W({\bf F},{\bf f}) $ is changed by inhomogeneity. 
Consequently, the values of the mean life of a state, 
the first moment of $ {\bf f}$ and the dynamical friction force
are changed by inhomogeneity with respect to those of homogeneous systems.\\
In this paper I show how the $W({\bf F})$ and  
$ W({\bf F},{\bf f}) $ distributions are changed by inhomogeneity and the effect of this change on dynamical friction.

The plan of the paper is the following: in Sec. ~2 I show how inhomogeneity changes $W({\bf F})$ and what is the effect of
having a finite number of bodies in the system.
In Sec. ~3, I calculate $ W({\bf F},{\bf f}) $ and its first moment in inhomogeneous systems.
In Sec. 4, I show the effect of inhomogeneity on dynamical friction. In Sec. 5, I check the theoretical results with N-body simulations and Sec. 6 is devoted to conclusions.

\section{The distribution function W(F) in inhomogeneous systems}

The force per unit mass acting on a test star of a finite N-body gravitating   
system is given by the usual formula: 
\begin{equation}
{\bf F}_{tot}( {\bf r})= 
-G\sum_{i=1}^{N} \frac{M_{i}}{|{\bf r}_{i}-{\bf r}|^{3}}
({\bf r}-{\bf r}_{i})
\label{eq:newton}
\end{equation}
where the sum is extended to the N stars, $M_{i} $ is the mass of 
the i-th field star and $ r_{i} $ is its distance relative to the origin. 
The value of $ {\bf F} $ at a given time depends on the 
instantaneous positions of all the other stars and therefore is subject 
to fluctuations as these position change.  
Even if the stellar distribution were constant and homogeneous the total 
force experienced by a star, $ {\bf F}_{tot}$, would fluctuate around an average value due to local Poisson 
fluctuations in the number density of neighboring stars.  
Under conditions typically found in astrophysical systems like
globular clusters and clusters of galaxies,  
the average force $ {\bf F}_{tot} $, acting 
on a test  star can be naturally decomposed as:
\begin{equation} 
{\bf F}_{tot} = {\bf F}_{pot}+{\bf F}_{stoch} \label{eq:tor}
\end{equation}
The first term on the right hand side of Eq.~\ref{eq:tor}
is the mean field force 
produced by the smoothed distribution of stars in the 
system and  
can be obtained from a potential function, $ \Phi({\bf r})$, which is also 
connected, by 
the Poisson equation, to the system mass density, $\rho({\bf r}) $.  
In the case of a truncated system having a power-law 
density profile:  
\begin{equation}
\rho(r) =\rho_{0} \left(\frac{r_{0}}{r}\right)^{p}, \hspace*{1cm}  0\leq r\leq R
\end{equation}
the mean internal gravitational force is given by: 
\begin{equation}
F_{pot}= -\frac{G M_{tot}}{R^{3-p}} r^{1-p}
\end{equation}
where $ G$ is the Gravitational constant, $ M_{tot}$ is the total 
mass of the system and $ R$ is its radius.
The second 
term on the right hand side of equation Eq.~\ref{eq:tor} is 
a stochastic variable describing the effects of the fluctuating 
part of the gravitational field. This component arises 
because of the discreteness of the mass distribution: it would be null in a continuous
system.
The fluctuating part of $ {\bf F}_{tot}$ 
that I have previously indicated $ {\bf F}_{stoch}$ can be studied by 
probabilistic methods defining a probability distribution of stochastic 
force $ W({\bf F}_{stoch}) $.  
In the case of an infinite homogeneous system, under 
the other hypotheses given in Sect. 1,   
the random force distribution law is given by  Holtsmark's distribution: 
\begin{equation}
W(F_{stoch})= \frac{2 F_{stoch}}{\pi} \int_{0}^{\infty} t d 
t \sin(t F_{stoch}) 
\exp\left[-n (Gm t)^{3/2} \frac{4 (2 \pi)^{3/2}}{15}\right] \label{eq:hol}
\end{equation}
(Chandrasekhar \& von Neumann 1942),  
where $ m $ is the mass of a field star, $ n $ the mean density  
and G the gravitational constant. This equation shows that the stochastic 
force probability distribution depends only on the mean density $ n$ or, 
equivalently, on the static configuration of the system. \\
The stochastic 
force probability distribution in inhomogeneous systems 
can be calculated in the same way and under the same hypotheses as that 
in homogeneous systems.   
Kandrup (1980) 
obtained an equation for the distribution 
of random force of an infinite inhomogeneous system with  
probability density  
$ \tau(r)= \frac{a}{r^{p}}$. 
The resulting equation is a generalization of the Holtsmark 
law:
\begin{equation}
W(F_{stoch})= \frac{2 F_{stoch}}{\pi} \int_{0}^{\infty} t d 
t \sin(t F_{stoch}) 
\exp \left[-\frac{\alpha}{2}(Gm t)^{(3-p)/2} \int_{0}^{\infty}
\frac{dz (z-\sin z)}{z^{(7-p)/2}}\right] 
\label{eq:holt}
\end{equation}
where $ \alpha= \frac{(3-p) N(R)}{R^{(3-p)}}=4\pi a$ (because the total number of stars
at a distance $R$ is given by: $N(R)=4\pi\int_{0}^{\infty}dr r^{2}\tau (r)=4\pi
aR^{(3-p)}/(3-p)$). Equation \ref{eq:holt} 
reduces to the Holtsmark distribution for a uniform system in the case $ p= 0$. 
The probability 
density must be chosen in such a way to ensure a convergent
integral in Eq.~\ref{eq:holt}: 
this restricts the choice to $ p <3 $. 
If the system is finite, as e.g is the case for globular clusters, the 
Holtsmark and Kandrup laws must be substituted by a law in which N is finite.
This last can be calculated as follows:
%
%
suppose to have a cluster of radius R 
containing N stars of mass $ m $
with probability distribution law given by $ \tau(r) = \frac{a}{r^{p}}$. 
I define the characteristic function of the probability 
distribution $ W({\bf F})$ \footnote{Note that from now on, $F$ stands for $F_{\rm stoch} $}:
\begin{equation}
C({\bf t})= 
\int \exp(i {\bf t F}) W({\bf F}) d {\bf F} \label{eq:car}
\end{equation}  
Suppose now that the system contains only one star.  
If $ W( {\bf F}_{i})$ is the probability distribution of the 
force due to the star calculated at the origin one verifies that:  
\begin{equation}
W({\bf F}_{i}) d {\bf F}_{i} = \tau({\bf r}) d {\bf r}
\end{equation}
Using the usual equation $ {\bf F}_{i} = \frac{ G m}{r_{i}^{3}} {\bf r}_{i}$
I obtain the probability distribution for one star:   
\begin{equation}
W_{1} ({\bf F}_{i})= \frac{a (Gm t)^{(3-p)/2}}{2 |{\bf F}_{i}|^{(9-p)/2}} 
\hspace{0.5cm} \frac{GM}{R^{2}}< F_{i}< \infty 
\label{eq:castt}
\end{equation}
Introducing Eq.~ \ref{eq:castt} into Eq.~\ref{eq:car}  the characteristic 
function for the force due to one star turns out to be given by: 
\begin{equation}
M({\bf t})= C_{1} ( {\bf t})= 
\frac{(3-p)(Gm t)^{(3-p)/2}}{2 R^{3-p}} \int_{\frac{G m t}{R^{2}}}^{\infty} 
\frac{\sin z }{z^{(7-p)/2}} d z 
\end{equation}
where $ z= | {\bf t} \cdot {\bf F}|$. Finally the probability 
distribution for $ {\bf F} $ is: 
\begin{equation}
W_{N} ({\bf F}) = \frac{1}{2 \pi^{2}} \int_{0}^{\infty} 
\frac{\sin(t |{\bf  F}|)}{ t |{\bf F}|} M^{N} ({\bf t})
t^{2} d t
\end{equation}
and the probability distribution of the modulus of the force is: 
\begin{equation}
W_{N} (F) = \frac{2 F}{\pi} \int_{0}^{\infty} t d t 
\sin(t F) \left[ \frac{(3-p)(Gm t)^{(3-p)/2}}{2 R^{(3-p)}} 
\int_{\frac{G m t}{ R^{2}}}^{\infty} 
\frac{\sin z}{z^{(7-p)/2}}d z\right]^{N} 
\label{eq:holter}
\end{equation}
that reduces to Holtsmark's law for $ N \rightarrow \infty$. 
In terms of $ \beta = \frac{F}{G m \alpha^{2/(3-p)}}$ and 
$ y = t G m \alpha^{\frac{2}{3-p}} $,  
Eq.~\ref{eq:holter} can be written as: 
\begin{equation}
W_{N} (\beta) = \frac{2 \beta}{\pi} \int_{0}^{\infty} y d y 
\sin(\beta  y) \left[ \frac{y^{(3-p)/2}}{2 N} 
\int_{\frac{y }{(3-p) N}}^{\infty} 
\frac{\sin z}{z^{(7-p)/2}}d z\right]^{N} 
\end{equation}
%
%
In the case the system is also clustered the $W_{\rm N}(F)$ distribution can be calculated 
following  Antonuccio-Delogu \& Atrio-Barandela (1992) (AA92).
I suppose that particles have the density 
distribution:
\begin{equation}
\tau(r) = \frac{a}{r^{p}} \exp\left(-\frac{r^{2}}{r_{0}^{2}}\right)
\end{equation}
where $a$ and $ p$ and $ r_{0}$ are three constants. 
The stochastic force distribution 
is given by:
\begin{equation}
W_{N}({\bf F}) = \int \frac{d^{3} t}{(2 \pi)^{3}} M_{N} ({\bf t}) 
\exp(- i {\bf t F}) \label{eq:step}
\end{equation}
The term $ M_{N}$ is given by: 
\begin{equation}
M_{N} ( {\bf t}) = \frac{A_{n}( {\bf t})}{N^{N}} 
\left[1+\frac{1}{2}(1-\frac{1}{N})\frac{\Sigma( {\bf t})}{A_{2} ({\bf t})}
\right]
\end{equation}
(AA92), where $ A_{n} ({\bf t})$ is given by:
\begin{equation}
A_{n} ({\bf t})=\left[ \frac{\alpha}{N} \frac{(G m t)^{(3-p)/2}}{2} 
\int_{\frac{Gm t}{R^{2}}}^{\infty} dz \frac{\sin z}{z^{ (7-p)/2}} 
\right]^{N} \label{eq:tost}
\end{equation}
$ A_{2} ({\bf t})$ is given by  
Eq.~\ref{eq:tost} with $ n=2 $ while 
$ \Sigma({\bf t})$ is given in the quoted paper (Eq.~34) and 
$ \alpha = \frac{ N}{ 2 \pi r_{0}^{3-p} \Gamma[(3-p)/2]}$.    
Integrating Eq.~\ref{eq:step} I finally obtain: 
\begin{eqnarray}
W_{N}(F)&=& 4 \pi^{2} |{\bf F}|^{2} W_{N} ({\bf F}) = 
\frac{2 F}{ \pi} \int_{0}^{\infty} t \sin( t F) 
\left[ \frac{\alpha}{N} (  G m t )^{ (3-p)/2} 
\int_{\frac{G m t }{ R^{2}}}^{\infty} \exp( -\frac{G m t}{ r_{0}^{2} z})
\frac{ \sin z }{z^{(7-p)/2}}\right]^{N} \nonumber\\
& &
\left[1 + \frac{1}{2}( 1- \frac{1}{N}) \frac{\Sigma(t)}{ A_{2} (t)}
\right]
\end{eqnarray}

\section{The distribution function W(F,f) in inhomogeneous systems} 

To calculate $ W({\bf F},{\bf f})$ in an 
inhomogeneous system I consider a particle moving with a velocity ${\bf v}$, 
subject to a force, per unit mass, given by 
Eq. (\ref{eq:newton})
and to a rate of change given by
\begin{equation}
{\bf f} = \frac{d{\bf F}}{dt} = G \sum_{i=1}^N M_i \left[ \frac{{\bf V}_i}{|{\bf r}_{i}|^{3}} - \frac{3 {\bf r}_{i} ({\bf r}_{i}{\bf V}_i)}{|{\bf r}_{i}|^{5}} \right]
\label{eq:sei}
\end{equation}
where $ {\bf V}_{i}$ is the velocity of the field particle relative to 
the test one. \\
The expression of $ W({\bf F},{\bf f})$ is given following Markoff's
method by (CN43):
\begin{equation}
W({\bf F},{\bf f}) = \frac{1}{64 \pi^{6}} \int_{0}^{\infty}
\int_{0}^{\infty}  A({\bf k},{\bf \Sigma})
\left\{ \exp{[-i({\bf k}{\bf \Phi} +
{\bf \Sigma}{\bf \Psi})]} \right\} d{\bf k} d{\bf \Sigma}
\label{eq:set}
\end{equation}
with $ A({\bf k}, {\bf \Sigma})$ given by 
\begin{equation}
A({\bf k}, {\bf \Sigma}) = e^{-nC({\bf k}, {\bf \Sigma})} \label{eq:setbis}
\end{equation}
being 
\begin{equation}
C({\bf k}, {\bf \Sigma}) = \int_{0}^{\infty}
\int_{- \infty}^{\infty}
\int_{- \infty}^{\infty} \tau \left[ 1 - \exp{i({\bf k}{\bf \Phi}
+ {\bf \Sigma}{\bf \Psi})} \right] d{\bf r} d{\bf V} dM
\label{eq:ot}
\end{equation}
where $ n$ is the average number of stars per unit volume while $ {\bf \Phi}$ 
and $ {\bf \Psi}$ are given by the following relations: 
\begin{equation} 
{\bf \Phi} = G \frac{M}{|{\bf r} |^{3}} {\bf r} 
\label{eq:nov}
\end{equation}
\begin{equation} 
{\bf \Psi} = \frac{d{\bf F}}{dt} = M \left[ \frac{{\bf V}}{|{\bf r} |^{3}} - \frac{3 {\bf r} ({\bf r} {\bf V})}{|{\bf r} |^{5}} \right]
\label{eq:die}
\end{equation}
and $ \tau ({\bf V},{\bf r},M) d {\bf V} d {\bf r}dM $ is
the probability that a star has velocity in the
range $ {\bf V}, {\bf V} + d{\bf V}$,  $ {\bf r}, {\bf r} + d{\bf r}$
and mass in $ M, M + dM$.\\
Now I suppose that 
$\tau$ is given by:
\begin{equation}
\tau = \frac{a}{r^p} \psi (j^2(M) |{\bf u}|^2)
\label{eq:un}
\end{equation}
where $ a$ is a constant
that can be obtained from the normalization condition for  $\tau$, $j$ a
parameter (of dimensions of velocity$^{-1}$), $\psi$ an arbitrary
function, $\bf u$ the velocity of a field star.
In other words I assume, according to CN43 and Chandrasekhar
\& von Newmann (1942), that the distribution of velocities is spherical,
i.e. the distribution function is 
$\psi({\bf u}) \equiv  \psi (j^2(M) |{\bf u}|^2) $,
but differently from the quoted papers I suppose that the positions
are not equally likely for stars, that is the stars are
inhomogeneously distributed in space.
Following Chandrasekhar, a lengthy calculation leads  
to find the function $A({\bf k}, {\bf \Sigma})$:
%
\begin{eqnarray}
A({\bf k}, {\bf \Sigma}) & = &= 
e^{ [- \; \;  \frac{\alpha}{2} ( G M k )^{\frac{3 - p}{2}} \cdot 
B(p) \; + \; \frac{i \alpha}{2} k^{\frac{- p}{2}} ( G M )^{\frac{2 - p}{2}}} \cdot  \nonumber \\
                         &   & \left| {\bf v} \right| \cdot [ \Sigma_{1}  \sin{\gamma} \; - \; 2 \Sigma_{3} \cos{\gamma} ] \cdot A(p) \; - \; \frac{\alpha}{8} ( G M )^{\frac{1 - p}{2}} k^{\frac{ -(3 + p)}{2}} \cdot \nonumber \\
                         &   &  \Gamma \left( \frac{p}{2} + \frac{3}{2} \right)  \cdot \left\{ \cos{\left[ - \frac{\pi}{4} \left( \frac{p}{3} + 1 \right) \right]} \; - \; \sin{\left[ - \frac{\pi}{4} \cdot \left( \frac{p}{3} + 1 \right) \right]}  \right
\} \cdot \nonumber \\ 
                         &   & \left\{ \left| {\bf u} \right|^{2} \cdot \left[ \frac{1}{3} \; ( a + b + c ) ( \Sigma_{1}^{2} +  \Sigma_{2}^{2} ) \; + \;  \frac{\Sigma_{3}^{2}}{3} 
( 2 c + d ) \right] \right. \nonumber \\
                         &   & + \; \left| {\bf v} \right|^{2} \cdot \left[ (a \sin^{2}{\gamma} + c \cos^{2}{\gamma}) \Sigma_{1}^{2} \; + \; (b \sin^{2}{\gamma} + c \cos^{2}{\gamma}) \Sigma_{2}^{2} \right. \nonumber \\
                         &   & + \; ( c \sin^{2}{\gamma} + d \cos^{2}{\gamma}) \Sigma_{3}^{2} \; + \; f \Sigma_{1} \Sigma_{3} 
\sin{\gamma} \cos{\gamma} \left] \left\} \right. \right.]
\label{eq:trenci}
\end{eqnarray}
where
\begin{equation}
k =  \frac{| {\bf v} |^{2}}{| {\bf u} |^{2}} \nonumber
\end{equation}
$ {\bf u}$ is the distribution of the velocities of the field stars, and $ {\bf v}$ is the velocity of the test star,
\begin{equation}
A(p) =  \int_{0}^{\infty} \left[  \frac{\sin{x}}{x^{(4 - p)/2}} -  \frac{3 \sin{x}}{x^{(8 - p)/2}} + \frac{3 \cos{x}}{x^{(6 - p)/2}} \right] \cdot d x  
\label{eq:venqua}
\end{equation}
\begin{equation}
B(p) = \int_{0}^{\infty} \frac{z - \sin{z}}{z^{(7 - p)/2}} \cdot d z
\end{equation}
\begin{equation}
a=  \frac{- 2}{p + 1} - \frac{ 24}{(p + 1) (p - 3) } - \frac{ 216}{(p + 1) (p - 3) (p - 7) } \nonumber
\end{equation}
\begin{equation}
b =  \frac{- 72}{(p + 1) (p - 3) (p - 7) } \nonumber
\end{equation}
\begin{equation}
c =  \frac{- 36}{(p - 3) (p - 7) } 
\label{eq:tren}
\end{equation}
\begin{equation}
d =    \frac{- 2}{p + 1} -  \frac{12}{3 - p} + \frac{18}{7 - p}  \nonumber
\end{equation}
\begin{equation}
e =  \frac{- 4}{p + 1} - \frac{ 48}{(p + 1) (p - 3) } - \frac{ 288}{(p + 1) (p - 3) (p - 7) } \nonumber
\end{equation}
\begin{equation}
f = \frac{- 4}{p + 1} - \frac{ 24}{(p + 1) (p - 3) } - \frac{12}{3 - p} + \frac{144}{(p - 3) (p - 7) } \nonumber
\end{equation}
$\gamma$ is the angle between $ {\bf k}$ and $ {\bf v}$, ${\bf \Sigma}=(\Sigma_1,\Sigma_2,\Sigma_3)$.
Eq. (\ref{eq:trenci}) introduced into Eq. (\ref{eq:set}) solves the problem
of finding the distribution $W(\bf F,\bf f)$ and makes it possible to find
the moments of $\bf f$ that give information regarding the dynamical
friction.
Eq. (\ref{eq:trenci}) can be written in a more compact form, namely: 
\begin{eqnarray}
A({\bf k}, {\bf \Sigma}) & = & e^{  -\tilde{a} k^{\frac{3-p}{2}}} \{1- \; i g p({\bf k}, {\bf \Sigma})   \nonumber \\
                            &   & + \; \tilde{b} k^{\frac{- (3 
+ p)}{2}} ] \cdot  [ Q({\bf \Sigma}) + k R({\bf \Sigma}) ] \} 
\label{eq:trense}
\end{eqnarray}
if we define:
\begin{equation}
\tilde{a} =  \frac{\alpha}{2} ( G M )^{\frac{3 - p}{2}} \cdot B(p) \nonumber
\end{equation}
\begin{equation}
g =   \frac{\alpha}{2} ( G M )^{\frac{2 - p}{2}}
\cdot  | {\bf v} | \cdot A(p) \nonumber
\label{eq:cas}
\end{equation}
\begin{equation}
p({\bf k}, {\bf \Sigma}) =  k^{\frac{-p}{2}}
\cdot [ \Sigma_1 \sin{\gamma} \; - \; 2 \Sigma_3 \cos{\gamma} ] \nonumber
\label{eq:cass}
\end{equation}
\begin{eqnarray}
\tilde{b} & = & \frac{\alpha}{8} ( G M )^{\frac{1 - p}{2}} \cdot \Gamma \left( \frac{p}{2} +  \frac{3}{2} \right) \; \cdot \nonumber \\ 
          &   & \left\{ \cos{ \left[ - \frac{\pi}{4} \left( \frac{p}{3} + 1 \right) \right]} - \sin{ \left[ - \frac{\pi}{4} \cdot \left( \frac{p}{3} + 1 \right) \right] } \right\} \cdot  | {\bf u} |^{2}  
\end{eqnarray}
\begin{equation}
Q({\bf \Sigma}) =  \left[ \frac{1}{3} 
\; ( 
a+ b + c ) ( \Sigma_{1}^{2} +  \Sigma_{2}^{2} )+
\frac{\Sigma_{3}^{2}}{3}
( 2 c + d ) \right] \nonumber
\end{equation}
\begin{eqnarray}
R({\bf \Sigma}) & = & (
a \sin^{2}{\gamma} + c \cos^{2}{\gamma}) \Sigma_{1}^{2} \; + \; (b \sin^{2}{\gamma} + c \cos^{2}{\gamma}) \Sigma_{2}^{2} \nonumber \\
                &   & + \; ( c \sin^{2}{\gamma} + d \cos^{2}{\gamma}) \Sigma_{3}^{2} \; + \; f \Sigma_{1} \Sigma_{3} 
\sin{\gamma} \cos{\gamma} \nonumber
\end{eqnarray}

As I stressed in the introduction, the study of the dynamical friction 
is possible when we know the first moment of $ {\bf f}$. This calculation 
can be done using the components of $ {\bf f}$ ($ f_i, f_j, f_k$). 
We have that:
\begin{equation}
f_i =  \frac{\int_{- \infty}^{\infty} W({\bf F},{\bf f})  f_i d{\bf f}}{W({\bf F})}
\label{eq:trenset}
\end{equation} 
The distribution 
function $ W({\bf F})$, giving the number of stars subject to a force 
$ {\bf F}$, can be calculated as follows:
\begin{equation}
W({\bf F}) =  \frac{1}{64 \pi^{6}} \int_{0}^{\infty} \int_{0}^{\infty} \int_{0}^{\infty} \{ e^{[-i({\bf k}{\bf F} + {\bf \Sigma}{\bf f})]} \} \cdot A({\bf k},{\bf \Sigma}) \; d {\bf k} d {\bf \Sigma} d {\bf f}
\label{eq:trenot}
\end{equation}
Integrating, I find:
\begin{equation}
W({\bf F}) =  \frac{1}{2 \pi^{2} F} \int_{0}^{\infty} \{ e^{ [ \; - a^2 k^{(3-p)/2} ] } \} \cdot  k \sin({k F}) dk
\label{eq:trenno}
\end{equation}
This equation gives the generalized Holtsmark distribution obtained by 
Kandrup (1980a - Eq. 4.17) and provides the probability that a star is 
subject to a force $ {\bf F}$ in a inhomogeneous system.\\
In order to calculate the first moment of $ {\bf f}$
we need only an approximated form for $ A({\bf k}, {\bf \Sigma})$ (see Chandrasekhar 1943):
\begin{equation}
A({\bf k}, {\bf \Sigma}) = \{ e^{[ \; - a^2 k^{(3-p)/2}]} \} \cdot [1 - i g p({\bf k}, {\bf \Sigma}) ] 
\label{eq:quar}
\end{equation}
Using this last expression for $A({\bf k}, {\bf \Sigma})$ and
Eq. (\ref{eq:cas}), Eq. (\ref{eq:cass}), Eq. (\ref{eq:trenset}), Eq. (\ref{eq:trenno}),
Eq. (\ref{eq:quar}) and performing a calculation similar to that by 
CN43 
the first moment of  $ {\bf f}$ is given by:
\begin{equation}
\overline {\bf f} =  -\left( \frac{1}{2} \right)^{\frac{3}{3 - p}} \cdot A(p) \cdot B(p)^{\frac{p}{3 - p}} \cdot \frac {\alpha^{\frac{3}{3 - p}} G M L(\beta)}{ \pi H(\beta) \beta^{\frac{2 - p}{2}}} \cdot \left[ {\bf v} - \frac{3 {\bf F} \cdot {\bf v}}{ |  {
\bf F} |^2} \cdot {\bf F} \right]
\label{eq:quaruno}
\end{equation}
where
\begin{eqnarray}
L ( \beta) & = & 6 \int_{0}^{\infty} \left[
e^{(x/\beta)^{\frac{(3 - p)}{2}}} \right]
\left[ \frac{ \sin{x}}{x^{(2 - p)/2}} -
\frac{\cos{x}}{x^{p/2}} \right] dx \nonumber \\
           &   & - \; 2 \int_{0}^{\infty}
           \left[ e^{(x/\beta)^{\frac{(3 - p)}{2}}}
           \right] \cdot \frac{\sin{x}}{x^{(p - 2)/2}} dx
\label{eq:quardu}
\end{eqnarray}
\begin{equation}
H(\beta) = \frac{2}{\pi \beta}
\int_{0}^{\infty} e^{-\left(\frac{x}{\beta}\right)^{3/2}}
\cdot x sin(x) dx
\end{equation}
The results obtained here for an inhomogeneous system 
are different [see Eq. (\ref{eq:quaruno})], as expected,
from those obtained by CN43 for
a homogeneous system (CN43 - Eq. 105 or Eq. (\ref{eq:quarot})). At the same time 
it is very interesting
to note that for $ p = 0$ (homogeneous system) the result here obtained coincides,
as is obvious, with the
results obtained by CN43. 
In fact, for $ p = 0$ Eq. (\ref{eq:quaruno}) reduces to:
\begin{equation}
\overline {\bf f} = - \frac{\alpha}{6} G m \left[ \frac{L(\beta)}{\pi \beta H(\beta)} \right]_{p = 0} \cdot \left[ {\bf v} - \frac{3 {\bf F} \cdot {\bf v}}{ | {\bf F} |^2} \cdot {\bf F} \right]
\label{eq:quartre}
\end{equation}
Defining
\begin{equation}
H ( \beta) = \frac{2}{\pi \beta} \int_{0}^{\infty} \left\{ e^{ \left[ (\frac{ - \; x }{\beta})^{3/2}\right]}  \right\} \cdot  x \sin{x} dx
\label{eq:quarsei}
\end{equation}
we have that 
\begin{equation}
[ L ( \beta)]_{p = 0} = \int_{0}^{\infty}
e^{ \left[ - \; (x/\beta)^{3/2} \right] }
\left[ \frac{ 6 \sin{x}}{x} - 6 \cos{x} + 2 x \sin{x} \right] dx=
3 \pi \int_{0}^{\beta} H( \beta) d \beta - \pi \beta H( \beta)
\label{eq:quarqua}
\end{equation}
In this way we can write Eq. (\ref{eq:quartre}) as:
\begin{equation}
\overline {\bf f} = - \frac{2 \pi}{3} G m n \left[ \frac{3 \cdot \int_{0}^{\beta} H(\beta) d \beta }{ \beta \cdot H(\beta)} \; - \; 1 \right] \cdot \left[ {\bf v} - \frac{3 {\bf F} \cdot {\bf v}}{ | {\bf F} |^2} \cdot {\bf F} \right]
\label{eq:quarset}
\end{equation}
this last equation coincides with Eq. (105) of CN43.\\

In inhomogeneous systems, Chandrasekhar equation
\begin{equation}
\langle \frac{d |{\bf F}|}{dt} \rangle= \frac{4 \pi}{3} G M n B(\beta) \cdot \frac{{\bf v F}}{{\bf F}} \label{eq:cinqcin}
\end{equation}
can be written, using
Eq. (\ref{eq:quaruno}), as:
\begin{equation}
\langle \frac{d |{\bf F}|}{dt} \rangle=  2^{\frac{p}{p-3}} \cdot A(p) \cdot B(p)^{\frac{p}{3 - p}} \cdot \frac {\alpha^{\frac{3}{3 - p}} G M L(\beta)}{ \pi H(\beta) \beta^{\frac{2 - p}{2}}} \cdot \frac{{\bf v F}}{{\bf F}}
\label{eq:q1}
\end{equation}
In order to check the validity of the quoted relation (Eq. (\ref{eq:q1})), I have performed numerical experiments.
This was done by evolving 100000 points (stars) acting under their mutual gravitational attraction. From the evolved positions and velocities of the stars, $\langle \frac{d {\bf F}}{dt} \rangle$ was computed as a function of velocity and force, similarly to Ahmad \& Cohen (1974), and then compared with Eq. (\ref{eq:q1}) as I shall describe in the following. 

As previously quoted, the results obtained in the present paper for an inhomogeneus system 
are different [see Eq. (\ref{eq:quaruno})], as expected,
from that obtained by CN43 for
a homogeneous system (CN43 - Eq. 105). 
In a inhomogeneous system, in a similar 
way to what happens in a homogeneous system, $ {\bf f}$ depends
on $ {\bf v}$, $ {\bf F}$ and $ \theta$ (the angle between $ {\bf v}$ and 
$ {\bf F}$) while differently from homogeneous systems, $ {\bf f}$ 
is a function of the inhomogeneity parameter
$p$. The dependence of $ {\bf f}$ on $p$ is not only due to the
functions $A(p)$, $ B(p)$ and to the density parameter $ \alpha$ but
also to the parameter $ \beta=|\bf F|/Q_{H}$. In fact in
inhomogeneous systems the {\it normal} field $Q_{H}$ is given by
$Q_{H}=GM(\alpha B(p)/2)^{2/(3-p)}$, clearly dependent on $p$.

\section{Dynamical friction in inhomogeneous systems}

The introduction of the notion of dynamical friction is due to CN43. In the 
stochastic formalism developed by CN43 the dynamical friction is discussed 
in terms of $ {\bf f}$:
\begin{equation}
\overline {\bf f} =  \frac{-2 \pi}{3} G m n  B(\beta) \left[ {\bf v} - \frac{3 {\bf F} \cdot {\bf v}}{ | {\bf F} |^2} \cdot {\bf F} \right]
\label{eq:quarot}
\end{equation}
where
\begin{equation}
B( \beta) =  \frac{3 \cdot \int_{0}^{\beta} W(\beta) d \beta }{ \beta \cdot W(\beta)} \; - \; 1
\label{eq:quarno}
\end{equation}
As shown by CN43, the origin of dynamical friction is due to 
the asymmetry in the distribution of relative velocities. If a test star moves with velocity $ {\bf v}$ in a spherical 
distribution of field stars, namely $ \phi( \bf u)$ then we have that:
\begin{equation}
\overline{\bf V} = \overline{{\bf u} - {\bf v}} = - {\bf v}
\label{eq:cinqqua}
\end{equation}
The asymmetry in the distribution of relative velocities is conserved 
in the final Eq. (\ref{eq:quarot}). In fact from Eq. (\ref{eq:quarot}) 
we have:
\begin{equation}
\frac{d |{\bf F}|}{dt} = \frac{4 \pi}{3} G M n B(\beta) \cdot \frac{{\bf v F}}{{\bf F}} \label{eq:cinqcin}
\end{equation}
(CN43). This means that when $ {\bf v} \cdot {\bf F} \ge 0$ then 
$ \frac{d |{\bf F}|}{dt} \ge 0$; 
while when $ {\bf v} \cdot {\bf F} \le 0$ then $ \frac{d |{\bf F}|}{dt} \le 0$.
As a consequence, when ${\bf F}$ has a positive component in the direction of
${\bf v}$,  ${|\bf F|}$ increases on average; while if  ${\bf F}$ has
a negative component in the direction of  ${\bf v}$,  ${|\bf F|}$
decreases on average. 
Moreover, the star suffers a greater amount of acceleration in the
direction $ - {\bf v}$ when
$ {\bf v} \cdot {\bf F} \le 0$ than in the direction $ +{\bf v}$ when 
$ {\bf v} \cdot {\bf F} \ge 0$.\\ 
In other
words the test star suffers, statistically, an equal number of 
accelerating and decelerating impulses. Being the modulus of 
deceleration larger than that of acceleration the star slows down.\\
At this point we may show how dynamical friction changes due to 
inhomogeneity. From Eq. (\ref{eq:quaruno}) we see that
$ \frac{d{\bf F}}{dt}$ differs from that obtained in homogeneous system 
only for the presence of a dependence on the inhomogeneity parameter 
$ p$. If we divide Eq. (\ref{eq:quaruno}) by Eq. (\ref{eq:quarot}) we 
obtain:
\begin{eqnarray}
\frac{ \left( \overline{\frac{d{\bf F}}{dt}} \right)_{Inh.}}
{ \left( \overline{\frac{d{\bf F}}{dt}} \right)_{Hom.}} 
& = & - \left( \frac{1}{2} \right)^{\frac{3}{3 - p}} \cdot A(p) \cdot B(p)^{\frac{p}{3 - p}} \cdot \frac{\alpha^{\frac{3}{3 - p}} L(\beta)}{ \pi H(\beta) \beta^{\frac{2 - p}{2}}} \cdot \frac{3}{2 \pi n B(\beta)} \; \nonumber \\
& = & \; -\frac{3}{n \cdot \pi} \cdot \left( \frac{1}{2} \right)^{\frac{6 - p}{3 - p}} \cdot A(p) \cdot B(p)^{\frac{p}{3 - p}} \cdot \frac{\alpha^{\frac{3}{3 - p}} L(\beta)}{ \pi H(\beta) B(\beta) \beta^{\frac{2 - p}{2}}}
\label{eq:cinqsei}
\end{eqnarray}

\begin{figure}
\psfig{figure=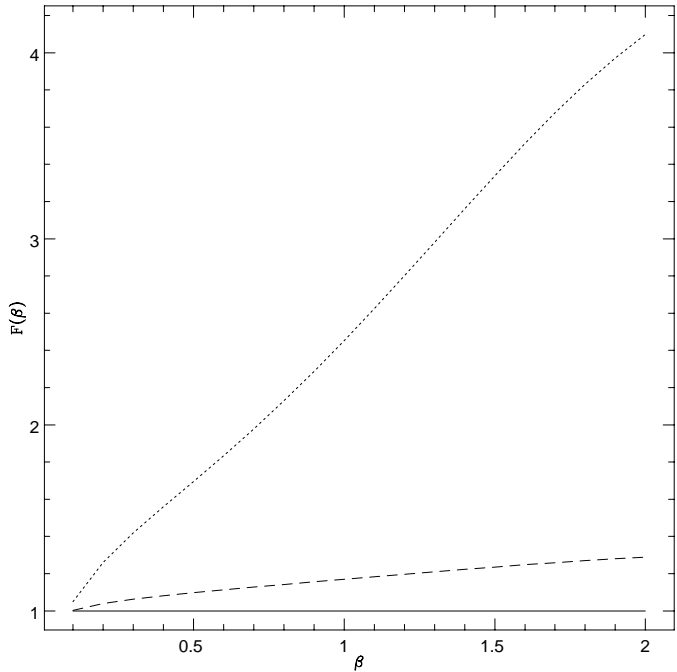,width=10cm} 
\caption[]{The function $F(\beta)$ for several values of the inhomogeneity 
parameter $p$; solid line $p=0$, dashed line $p=0.1$; dotted line $p=0.5$}
\end{figure}
If we consider a homogeneous system, $ p=0$, the previous equation
reduces to:
\begin{equation}
\frac{ \left( \overline{\frac{d{\bf F}}{dt}} \right)_{Inh.}}
{ \left( \overline{\frac{d{\bf F}}{dt}} \right)_{Hom.}} 
= 1
\end{equation}
In the case of an inhomogeneous system, $ p \neq 0$, we see that:
\begin{equation}
\frac{ \left( \overline{\frac{d{\bf F}}{dt}} \right)_{Inh.}}
{ \left( \overline{\frac{d{\bf F}}{dt}} \right)_{Hom.}} 
=n^{p/(3-p)} F(\beta(n,p))
\end{equation}
where
\begin{equation}
F(\beta(n,p) =  \; -\frac{3}{\pi} \cdot \left( \frac{1}{2} \right)^{\frac{6 - p}{3 - p}} \cdot A(p) \cdot B(p)^{\frac{p}{3 - p}} \cdot \frac{L(\beta)}{ \pi H(\beta) B(\beta) \beta^{\frac{2 - p}{2}}}
\end{equation}

\begin{figure}
\psfig{figure=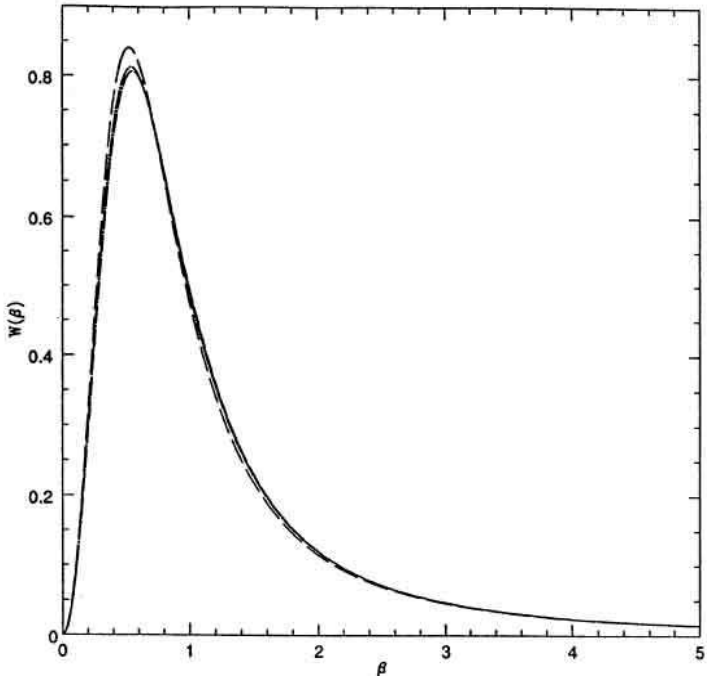,width=10cm} 
\caption[]{The theoretical probability distribution of stochastic force 
in an inhomogeneous system for different values of N. The solid 
line is Kandrup's distribution for an infinite inhomogeneous system, the 
long dashed line is Kandrup's distribution for a finite inhomogeneous 
system with $N= 50$, the dotted line is the same distribution 
for $ N=100$. The Kandrup's distribution for a finite inhomogeneous 
system with $ N = 1000$ is indistinguishable from that  
for the infinite system. The force is measured 
in units of $ G m \alpha^{2/(3-p)}$.
}
\end{figure}

\begin{figure}
\psfig{figure=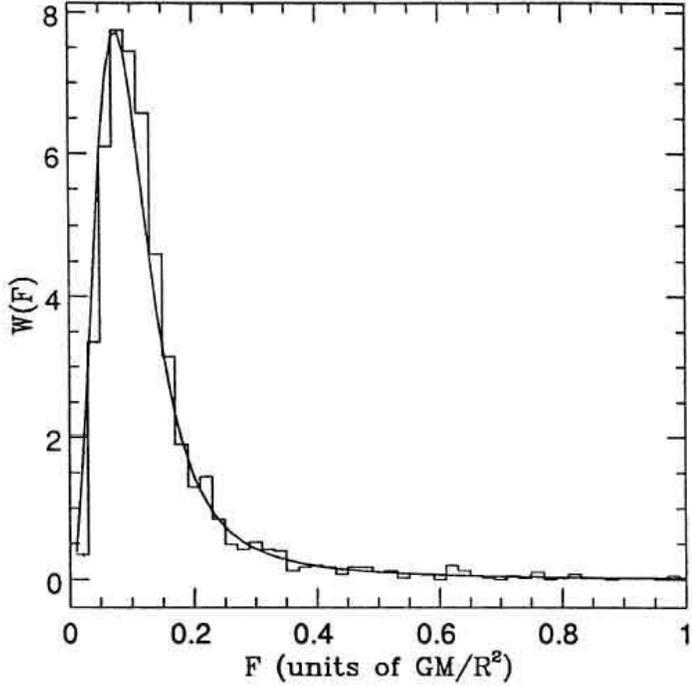,width=10cm} 
\caption[ ]{Experimental distribution of the stochastic force 
in an inhomogeneous system. The solid line is Kandrup's distribution 
for a finite inhomogeneous system ( $ p= 0.01 $) 
of 100000 particles, the histogram is the experimental distribution 
of stochastic force obtained from a system of 100000 particles 
as described in the text. The force is measured 
in units of $ \frac{G M}{R^{2}}$.}\end{figure}

\begin{figure}
\psfig{figure=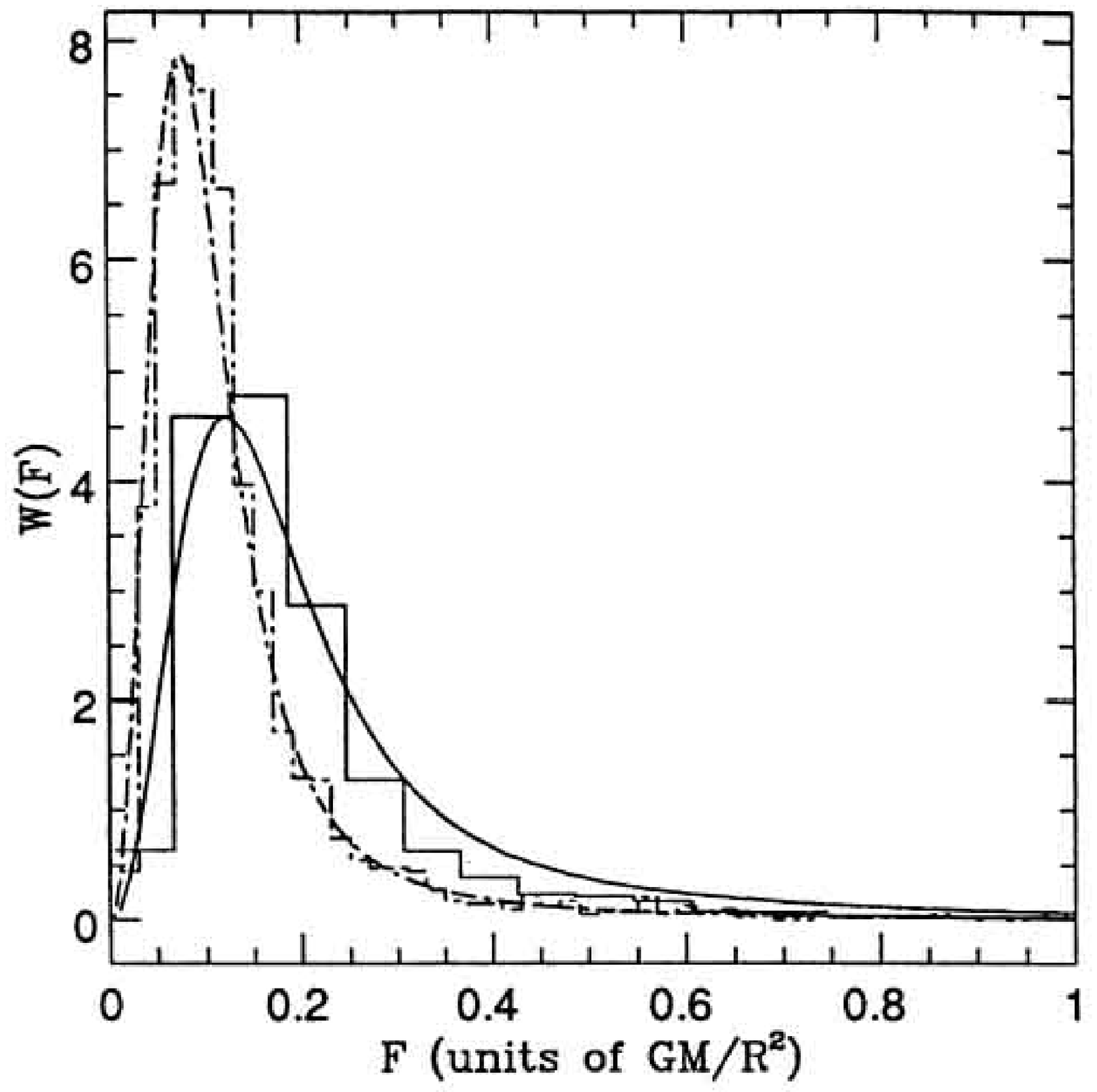,width=10cm} 
\caption[ ]{Experimental distribution of stochastic force in 
a clustered system. The solid line is the AA(92)  
distribution for 
a clustered system of 100000 particles. The histogram is the experimental 
distribution of stochastic force obtained from a clustered system of 
100000 particles as described in the text. The dot dashed line 
and histogram are a Holtsmark's distribution and an experimental 
distribution for an homogeneous 
system of 100000 particles.}
\end{figure}

As I show in Fig. 1, 
this last  equation is an increasing function of $p$. This means
that for increasing values of $ p$ 
the star suffers an even greater amount of acceleration in the
direction $ - {\bf v}$ when
$ {\bf v} \cdot {\bf F} \le 0$ than in the direction $ +{\bf v}$ when 
$ {\bf v} \cdot {\bf F} \ge 0$, with respect to the homogeneous case.
This is due to the fact that the difference
between the amplitude of the decelerating impulses and the accelerating ones 
is, as in homogeneous systems, statistically negative, but now larger,
being the scale factor greater.
This finally means that, for a given value of $n$, 
the dynamical friction increases with increasing inhomogeneity in
the space distribution of stars 
(it is interesting to note that this effect is fundamentally due to the
inhomogeneity of the distribution of the stars and not to the density
$n$). In other words two systems
having the same $n$ will have
their stars
slowed down differently according to the value of $p$. This is strictly
connected to the asymmetric origin of the dynamical friction. \\
In addition, by 
increasing $n$ the dynamical friction increases, just like in the 
homogeneous systems, but the increase is larger than the 
linear increase observed in homogeneous systems.

\section{N-body experiments}


To calculate the stochastic force in an inhomogeneous system, 
I used an initial configuration in which particles were 
distributed according 
to a truncated power-law 
density profile:  
\begin{equation}
\rho(r)=\rho_0 \left(\frac{r_0}{r}\right)^p  \hspace*{1cm}  0\leq r\leq R
\end{equation}
(see Kandrup 1980). 
If the velocity distribution is everywhere isotropic then the equation relating the 
configuration space density $\rho(r)$ to the phase space density f(E) is:
\begin{equation}
\rho(r)=4\pi \int_{U(r)}^{0} \sqrt{2\left[E-U(r)\right]} f(E) dE
\label{eq:confff}
\end{equation}
where $U(r)$ is the potential (normalized to zero at infinity). Eq. (\ref{eq:confff})
may be converted into an Abel integral equation and inverted, giving
the phase space density:
\begin{equation}
f(E)=\frac{\sqrt{2}}{4 \pi^2} \frac{d}{dE} \int_E^0 
\frac{dU}{\sqrt{U-E}}
\frac{d \rho}{dU}
\label{eq:abell}
\end{equation}
(Eddington 1916; Binney \& Tremaine 1987).
\begin{figure}
\label{Fig. 1}
\psfig{figure=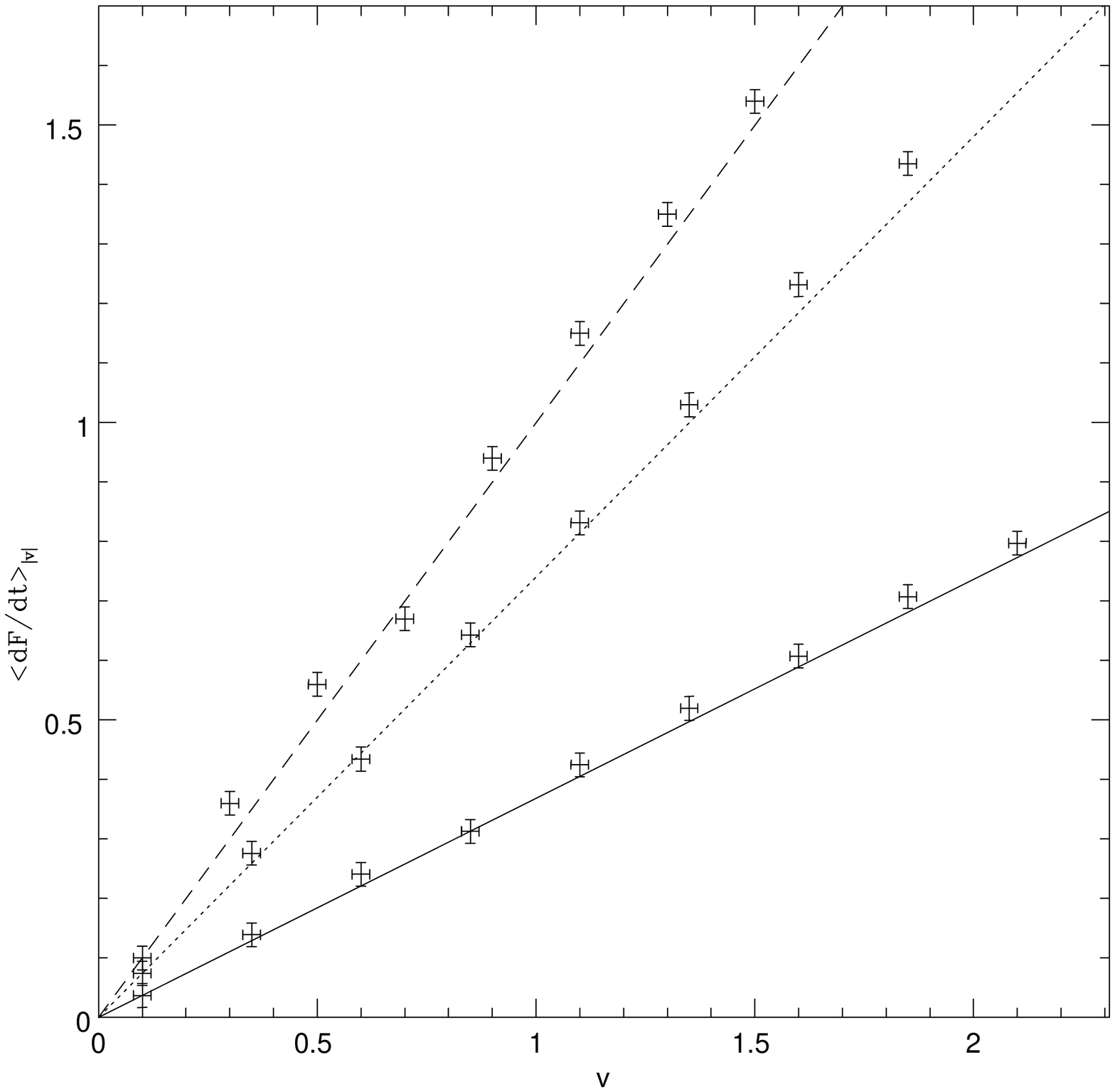,width=10cm} 
\caption[]{The average value of the time rate of change of the magnitude of the force as the function of the velocity.
The solid line refers to the homogeneous case (Chandrasekhar \& von Neumann 1943). The dotted and dashed line refers to the cases $p=2.5$ and $p=4$, respectively (see Eq. 26). Crosses represent the experimental points.}
\end{figure}
The initial conditions were generated 
from the distribution function that can be obtained from 
Eq.~(\ref{eq:abell})
assuming a cut-off radius $ R=1$, the mass of the system $ M=1$, 
$ r_{0} =0.15$ \footnote{This is the value I used, remember however that 
the distribution is scale-free} and $ G=1$. All the particles had  
equal mass.  
To have a system whose total mass is contained in a unitary 
sphere, Eq. (\ref{eq:confff})
was renormalized and 
consequently also the potential of the system 
which is obtained from Eq. (\ref{eq:confff})
through Poisson's equation. 
The system of 100000 particles 
was evolved over 150 dynamical times using a tree N-body code 
(Hernquist 1987). 
During the evolution of the system the essential quantities such as position, force, etc.,  
for each test point of the system was sampled 
every $ \frac{1}{20}$ of a dynamical time (see Ahamad \& Cohen 1973 for details). 
The stochastic force, $ {\bf F}_{stoch}$, was calculated observing 
that at the centre of a spherical system we have: 
\begin{equation}
{\bf F}_{tot} = {\bf F}_{stoch}+{\bf F}_{med} = {\bf F}_{stoch}
\end{equation}
because the mean field force, $ {\bf F}_{med}$, is equal to zero.
The force was calculated on a point at the centre of the system 
because theoretically Kandrup's distribution gives the probability 
distribution of the stochastic force for a particle at the centre only. 
When points displaced away from 
the centre are used the stochastic force distribution 
must be calculated as follows: \\
a) the stochastic force should be calculated subtracting the mean field 
force from the total force: 
\begin{equation}
{\bf F}_{stoch}={\bf F}_{tot}-{\bf F}_{med}
\end{equation}
b) the theoretical distribution must be numerically simulated as done by Ahmad 
\& Cohen (1973). 

\begin{figure}
\label{Fig. 1}
\psfig{figure=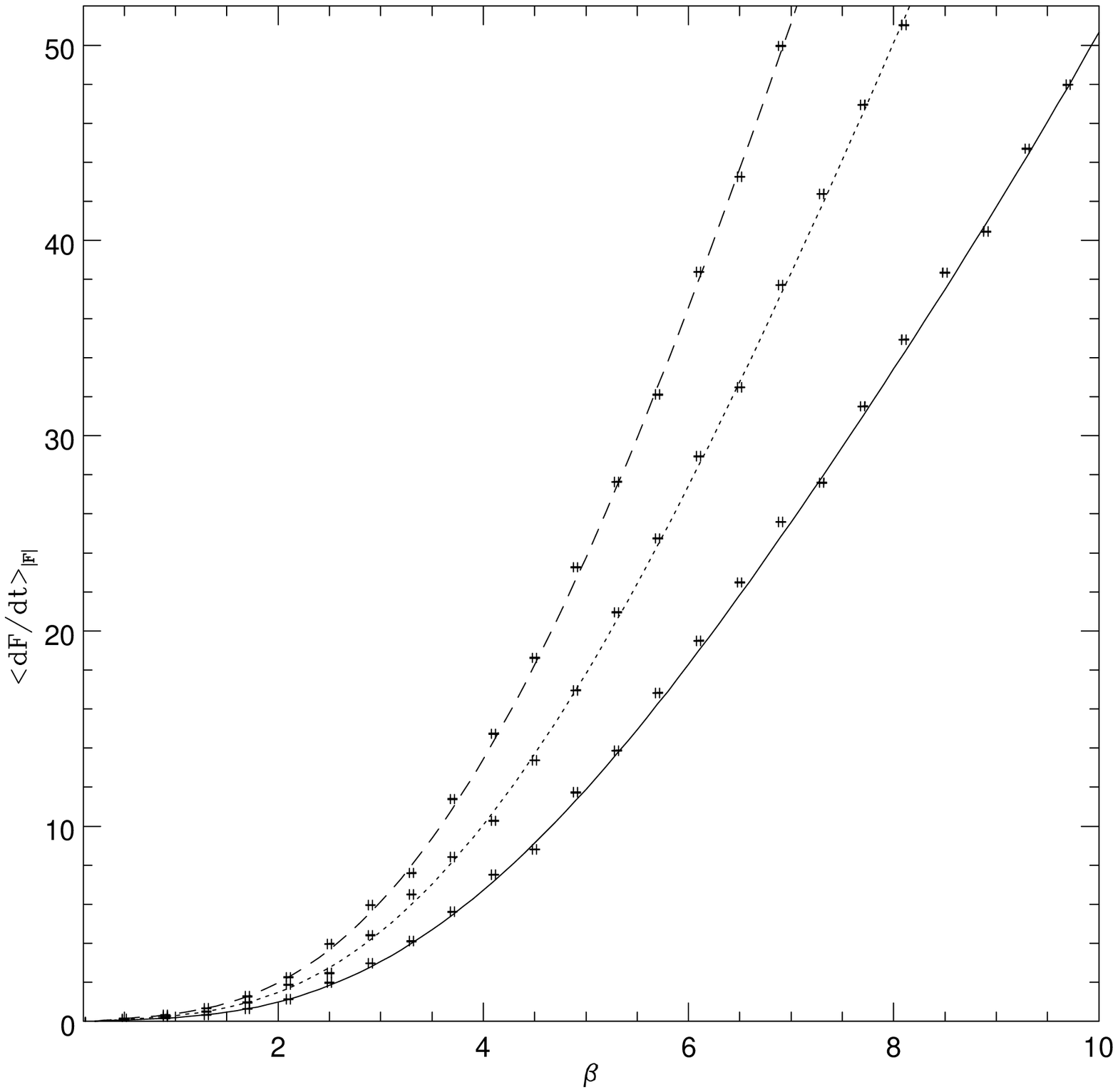,width=10cm} 
\caption[]{The average value of the time rate of change of the magnitude of the force as the function of the force.
The solid line refers to the homogeneous case (Chandrasekhar \& von Neumann 1943). The dotted and dashed line refers to the cases $p=2.5$ and $p=4$, respectively (see Eq. 19). Crosses represent the experimental points.}
\end{figure}
The average of the $ \frac{d |{\bf F}|}{dt}$ is a function of velocity, $v$, force, $F$, 
and the angle between them. The test of Eq. (\ref{eq:q1}) was performed in a similar way to that of 
Ahmad \& Cohen (1974), namely by integrating out two of the variables and examining
$ \langle \frac{d |{\bf F}|}{dt} \rangle$ against the remaining one (see Ahmad \& Cohen 1974).
As in Ahmad \& Cohen (1974), $ \langle \frac{d |{\bf F}|}{dt} \rangle_{|{\bf F}|}$ indicates $ \langle \frac{d |{\bf F}|}{dt} \rangle$ after integrating out the angle and velocity, while $ \langle \frac{d |{\bf F}|}{dt} \rangle_{|{\bf v}|}$ is 
$ \langle \frac{d |{\bf F}|}{dt} \rangle$ after integrating out the angle and force. In integrating out the angle, 
one cannot average over the entire range 0 to $\pi$, since that would give zero. 
Instead, the cosine was averaged from 0 to $\pi$/2, and in the numerical experiments only those particles having a cosine in the quoted range were used. In order not to waste the statistics for half the particles the same trick of Ahmad \& Cohen (1974) was used, namely when the cosine is in the range of $\pi/2$ to $\pi$, the sign of $ \frac{d |{\bf F}|}{dt}$ is changed and it is counted in the same statistics. This corresponds to assume that the cosine between ${\bf v}$ and ${\bf F}$ is uniformly distributed, which is what is found in numerical experiments. The average value of the cosine between 0 and $\pi/2$ is $1/2$. 

For a general distribution $\langle |{\bf v}| \rangle$ can be calculated as usual:
\begin{equation}
\langle |{\bf v}| \rangle=\int_0^{\infty}\frac{f(E) {\bf v} d{\bf v}}{f(E)d{\bf v}}
\end{equation}
and $\langle \frac{d |{\bf F}|}{dt} \rangle$
can be written, in units of $\frac{\left(\frac{1}{2}\right)^{\frac{3}{3-p}}}{\pi} \alpha^{\frac{3}{3 - p}} G M \langle |{\bf v}| \rangle$, as:
\begin{equation}
\langle \frac{d |{\bf F}|}{dt} \rangle_{|{\bf F}|}=
A(p) \cdot B(p)^{\frac{p}{3 - p}} \cdot \frac { L(\beta)}{ H(\beta) \beta^{\frac{2 - p}{2}}} \label{eq:q11}
\end{equation}
In the particular case of 
a Maxwellian distribution for velocities:
\begin{equation}
\psi= \frac{j^3}{\pi^{3/2}} exp(-j^2 |{\bf v}|^2)
\end{equation}
where $j^2=\frac{3}{2 <{\bf v}^2>}$, so that:
\begin{equation}
\langle |{\bf v}| \rangle=\frac{2}{\pi^{1/2}j}
\end{equation}
we have that:
\begin{equation}
\langle \frac{d |{\bf F}|}{dt} \rangle_{|{\bf F}|}=  \frac{2^{\frac{p}{p-3}}}{\pi^{3/2}j} \cdot A(p) \cdot B(p)^{\frac{p}{3 - p}} \cdot \frac {\alpha^{\frac{3}{3 - p}} G M L(\beta)}{ H(\beta) \beta^{\frac{2 - p}{2}}} 
\label{eq:q2}
\end{equation}
that expressed in units of 
$\frac{2^{\frac{p}{p-3}}}{\pi^{3/2}j} \alpha^{\frac{3}{3 - p}} G M$ 
, then Eq. (\ref{eq:q2}) becomes:
\begin{equation}
\langle \frac{d |{\bf F}|}{dt} \rangle_{|{\bf F}|}=  A(p) B(p)^{\frac{p}{3 - p}} \cdot \frac { L(\beta)}{ H(\beta) \beta^{\frac{2 - p}{2}}} 
\label{eq:q3}
\end{equation}
 
Similarly to Ahamd \& Cohen (1974), since to integrate out the force from Eq. (\ref{eq:q1}) one has a divergent result, I consider only particles up to a certain maximum value of the force, $\beta_{\rm max}$: for example in the case $p=0$,
$\beta_{\rm max}=10.7$, that involves 97\% of the particles. As observed by Ahmad \& Cohen (1974), any cutoff of the force can be used as long as it is taken into account in both the experiment and the analytic evaluation of $\langle \frac{d |{\bf F}|}{dt} \rangle_{|{\bf v}|}$. 
Defining:
\begin{equation}
B_1(\beta)=\frac { L(\beta)}{\pi H(\beta) \beta^{\frac{2 - p}{2}}} 
\end{equation}
and 
\begin{equation}
\langle B_1(\beta) \rangle=\frac{\int_0^{\beta_{\rm max}} B_1(\beta) H(\beta) d\beta}{\int_0^{\beta_{\rm max}} H(\beta) d\beta}
\label{eq:med}
\end{equation}
we find, in units of $\left(\frac{1}{2}\right)^{\frac{3}{p-3}} \langle B1(\beta)\rangle \alpha^{\frac{3}{3 - p}} G M$,
that:
%
\begin{equation}
\langle \frac{d |{\bf F}|}{dt} \rangle_{|{\bf v}|}=
A(p) \cdot B(p)^{\frac{p}{3 - p}} \cdot 
v
\label{eq:q5}
\end{equation}

The results of calculation and numerical experiments are plotted in Fig. 2-6.

As Ahmad \& Cohen(1973) showed, the stochastic force probability 
distribution for an infinite homogeneous system (Eq.~\ref{eq:hol}) 
and that for a finite one (Eq.~\ref{eq:holter} with $ p=0 $)
almost coincide for $ N \simeq 1000$. 
In Fig.~2, I show that the same result holds 
in the inhomogeneous case ($p = 0.5$). 
The different curves are obtained from 
Eq.~\ref{eq:holter}, describing the stochastic 
force distribution in a finite inhomogeneous system (long-dashed line), for increasing 
value of $ N $ ($ N = 50, 100, 1000$) and from Eq.~\ref{eq:holt}, 
which gives the stochastic force distribution in an infinite inhomogeneous 
system (solid line). When $ N \simeq 1000$ the two distributions are 
indistinguishable, meaning that for $ N \simeq 1000$ the 
stochastic force in an inhomogeneous system 
are equivalently described by the Kandrup's law for an infinite 
(Eq.~\ref{eq:holt}) or finite system (Eq.~\ref{eq:holter}).   
In Fig.~3 I compare Kandrup's  distribution for a 
finite inhomogeneous system ($ p=0.01$) 
with the histogram of forces obtained from a system of 100000 
particles. The plot shows a good agreement between the theoretical and numerical distribution.
In Fig.~ 4, I study the distribution of stochastic force in 
a clustered system. The solid line is the AA(92)  
distribution for a clustered system of 100000 particles. The histogram is the experimental 
distribution of stochastic force obtained from a clustered system of 
100000 particles. The dot dashed line and histogram are a Holtsmark's distribution and an experimental 
distribution for an homogeneous system of 100000 particles.
The plot shows as expected (Prigogine \& Severne 1966; Gilbert 1970; 
AA92) that there is an increase 
in higher force probability with respect to homogeneous 
and inhomogeneous systems. 
In Fig. 5, I plot the average value of the time rate of change of the magnitude of the force as a function of the velocity. The solid line refers to the homogeneous case while the dotted and dashed lines refer to the cases $p=2.5$ and $p=4$, respectively. Crosses represent the experimental points. 
As shown, experimental points follow a linear relationship and there is a good agreement with the theoretical prediction, (Eq. \ref{eq:q11}).
In Fig. 6, I plot the average value of the time rate of change of the magnitude of the force as the function of the force.
As in the previous figure, the solid line refers to the homogeneous case while the dotted and dashed line refers to the cases $p=2.5$ and $p=4$, respectively. In this case, the dependence is no longer linear: it behaves like $B(\beta)$ in the homogeneous case.
A comparison with numerical experiments shows that there is a good agreement with the theoretical prediction, (Eq. \ref{eq:q5}).
The situation described in Ahmad \& Cohen (1974), 
that the experimental data were somewhere in between the theoretical curves for the one-particle and the infinite-particle case, is no longer present and the agreement is better, now. This is due to the larger number of particles used in the simulations. The plots show that Chandrasekhar \& von Neumann's theory of dynamical friction in gravitational systems gives a good description of experimental data (solid line and data), and so does the generalization of the quoted theory to inhomogeneous systems (dotted line, dashed line, and respective data). In inhomogeneous systems, Chandrasekhar's result 
which relates the frictional force only to the local properties of the background at the position of the object, is no
longer true, and friction depends on the global structure of the system.  
This point is in agreement with Maoz (1993), who showed that in inhomogeneous media the friction, 
unlike Chandrasekhar's formula, depends on the global structure of the entire mass density field.

\section{Conclusions}

In this paper I showed how the distributions W(F), W(F,f) and 
the first moment $ {\bf f} = d {\bf F}/dt$ are changed in an inhomogeneous system. 
I obtained an expression for W(F) in finite inhomogeneous systems, an expression for 
W(F,f) and one relating $ {\bf f} = d {\bf F}/dt$ to 
the degree of inhomogeneity in a gravitational system. 
I showed the implications of this result on the dynamical friction in a
inhomogeneous system and in particular how
inhomogeneity acts as an amplifier of the asymmetry
effect giving rise to dynamical friction. 
Moreover, I tested by numerical simulations the previous results.
For what concerns $W(F)$, I showed that Kandrup's (1980) theory, describing the stochastic force 
probability distribution in infinite inhomogeneous systems, and AA92
theory, giving the stochastic force probability 
distribution in weakly clustered systems, describe correctly the observed
behavior. The results shows that in inhomogeneous and clustered systems there is an increase 
in higher force probability with respect to homogeneous systems. 
Furthermore I showed that for $ N> 1000$ Kandrup's 
theory can be applied to finite systems of particles.

In agreement with Ahmad \& Cohen (1974), 
the stochastic theory of dynamical friction developed by 
Chandrasekhar \& von Neumann (1943), in the case of homogeneous gravitational systems, gives a good description of the results of numerical experiments.
The stochastic force distribution obtained for inhomogeneous systems, in the present paper,
is also in good agreement with the results of numerical experiments.
Finally, in an inhomogeneous
background the friction force is actually enhanced relative to the
homogeneous case.


\begin{thebibliography}{}
\bibitem {ah1} Ahmad A., Cohen L., 1973, ApJ, 179, 885
\bibitem {ah2} Ahmad A., Cohen L., 1974, ApJ, 188, 469 
\bibitem{ant} Antonuccio-Delogu, V., Atrio-Barandela, F., 1992, Ap.J. 392, 403
\bibitem{} Antonuccio-Delogu V., Colafrancesco S., 1994, ApJ, 427, 72   
\bibitem{bas} Bahcall N.A., Soneira R.M., 1982, ApJ, 262, 419 
\bibitem{} Bontekoe, T. R., van Albada, T. S., 1987, MNRAS, 224, 349
\bibitem{} Binney J., Tremaine S., 1987, Galactic Dynamics, in Princeton Series in Astrophysics,
Princeton University Press.
\bibitem{} Chandrasekar S., 1941, ApJ, 94, 511
\bibitem{} Chandrasekar S., 1943a, Rev. Mod. Phys., 15, 1
\bibitem{} Chandrasekar S., 1943b, ApJ, 97, 255
\bibitem{} Chandrasekar S., 1943c, ApJ, 97, 263
\bibitem{} Chandrasekar S., 1943d, ApJ, 98, 25
\bibitem{} Chandrasekar S., 1943e, ApJ, 98, 47
\bibitem{cha2} Chandrasekhar S., 1944a, ApJ, 99, 47
\bibitem{cha3} Chandrasekhar S., 1994b, ApJ, 99, 25
\bibitem{} Chandrasekhar S., von Neumann J., 1942, ApJ, 95, 489
\bibitem{} Chandrasekhar S., von Neumann J., 1943, ApJ, 97, 1 (CN43)
\bibitem{} Dominguez-Tenreiro, R., Gomez-Flechoso, M. A., 1998, MNRAS 294, 465
\bibitem{edd} Eddington, A.S., 1916, M.N.R.A.S., 76, 572
\bibitem{} Elson R., Hut P., Inagaki S., 1987, ARA\&A, 25, 565
\bibitem{gil} Gilbert, I., 1970, Ap.J., 159, 239
\bibitem{her} Hernquist, L., 1987, Ap.J. Supp. Ser. 64, 715 
\bibitem{} Holtsmark P.J., 1919, Phys. Z., 20, 162
\bibitem{kan} Kandrup H.E., 1980a, Phys. Rep., 63, n. 1, 1
\bibitem{kdr} Kandrup H.E., 1980b, ApJ, 244, 1039
\bibitem{kdr} Kandrup H.E., 1983, Ap.\& S.S., 97, 435
\bibitem{} Kashlinsky A., 1986, ApJ, 306, 374
\bibitem{} Kashlinsky A., 1987, ApJ, 312, 497
\bibitem{lid} Liddle A.R., Lyth D.H.,1993, Phys. Rep., 231, n 1, 2 
\bibitem{} Maoz E., 1993, MNRAS 263, 75
\bibitem{pee} Peebles P.J.E., 1980,  ``The large scale structure of the Universe", Priceton University Press, Princeton 
\bibitem{pri} Prigogine, I., Severne, G., 1966, Physica, 32, 1234
\bibitem{sar} Sarazin C., 1988,  ``X-ray emission from Clusters of Galaxies", (Cambridge: Cambridge Univ. Press)
\bibitem{} Seguin, P., Dupraz, C., 1996, A \& A, 310, 757
\bibitem{stw} Strauss M.A., Willick J.A., 1995, Phys. Rept., 261, 271
\bibitem{whi} White S.D.M., Briel U.G., Henry J.P., 1993, MNRAS, 261, L8
\bibitem{} Wybo M., Dejonghe H., 1996, accepted A \& A 295, 347 
\bibitem{} Zwart S.F.P., Tout C.A., Lee, H.M., 1998, HiA 11, 622
(Highlights of Astronomy Vol. 11, Kluwer Academic Publishers, ed. Johannes Andersen) 
\end{thebibliography}
\end{document}